# The Rule of link functions on Binomial Regression Model: A Cross Sectional Study on Child Malnutrition, Bangladesh

Md Mehedi Hasan Bhuiyan

Department of Statistics and Data Science, University of Central Florida, USA

**Abstract**

Link function is a key tool in the binomial regression model defined as non-linear model under GLM approach. It transforms the nonlinear regression to linear model with converting the interval $(-\infty, \infty)$ to the probability $[0,1]$. The binomial model with link functions (logit, probit, cloglog and cauchy) are applied on the proportional of child malnutrition age 0-5 years in each household level. Multiple Indicator Cluster survey (MICS)-2019, Bangladesh was conducted by a joint cooperation of UNICEF and BBS . The survey covered 64000 households using two stage stratified sampling technique, where around 21000 household have children age 0-5 years. We use bi-variate analysis to find the statistical association between response and sociodemographic features. In the binary regression model, probit model provides the best result based on the lowest standard error of covariates and goodness of fit test (deviance, AIC).

Keywords: Proportion of child Malnutrition Age 0-5 years, Binomial Regression Model, Link function, Bangladesh

## 1. Introduction:

As a developing country, child malnutrition is decreasing steadily over the decades in Bangladesh (stunted: 42% in 2013, 28% in 2019; wasted: 9.6% in 2013, 9.8% in 2019; underweight: 22% in 2013,16.6% in 2019 ; Overweight: 1.6% in 2013, 2.4% in 2019) [1, 2]. Still now, this is an alarming issue among the policy makers. Child malnutrition, deficiency of minerals, vitamins and micro-nutrient, are more vulnerable to be affected different chronic diseases: respiratory issue, lung problem, diarrhea etc. These children also experience weak immune system that has long term affect in future. Anthropometric measurements (age, height and weight) are used assessing nutritional status of children. Child under-nourishment is measured by stunted (height-for-age, $< -2SD$), wasted (weight-for-height, $< -2SD$), underweight (weight-for-age, $< -2SD$) and overweight (weight-for-height, $> 2SD$). In this study, proportion of child malnutrition in each household level is considered as target variable [3].

Classical linear regression model is defined as a linear relationship between predictors and numerical response variable. The fitted value or the expected value of the model is a finite numerical value belongs to $(-\infty, \infty)$. On the other hand, Generalized Linear Model (GLM) works on categorical or counting response variable. Under the GLM approach, binomial regression model is used when response is binary like yes or no; success or failure. This model is also used for the proportional case of success or failure. The binomial model provides the probability of success or failure of given predictors. Since GLM is a non-linear function. The link function in the GLM transforms non-linear form to a linear form. There are several link functions like logit, probit, cloglog,etc. These functions are commonly used in the binomial regression model, [4,5].

## 2. Methodology:

### 2.1. Data Collection:
Bangladesh Bureau of Statistics (BBS) conducted the Multiple Indicator Cluster Survey (MICS), 2019, executed by UNICEF. Two stage sampling method was used in the survey covering 64 districts with 64000 households and 3220 primary sampling unit (PSU). Probability proportional to size (PPS) sampling was used to select census enumeration area in each stratum. Systematically, twenty households were chosen from each PSU.

### 2.2. Variable Selection:
We want to figure out child malnutrition age 0-5 years in each household level. The malnutrition consists of stunted, wasted, underweight and overweight. Proportion of child malnutrition is dependent variable of the study, the ratio of sum of malnourished children in four categories in a household and total number of children under five year in the same household level.

Socio-demographic factors are define as predictor variables, Mother education level (non-educated, primary, secondary, higher), father education (non-educated, primary, secondary, higher), mother antenatal care (yes, no), gender (male, female), area (urban, rural), wealth index (poor, middle, rich), child weight at birth (low, average, high), delivery place (home, hospital).

**2.3. Method:**

If the response variables $Y_1, Y_2 ... Y_n$ are independent, proportion of success, $\pi_i = Y_i/n_i$, where $Y_i$ is total number of malnourished children age 0-5 years, and $n_i$ is total number of children in the household age 0-5 years and $Y_i \sim$ binomial $(n_i, \pi_i)$. The likelihood function is

$$l(\beta, y) = \sum_{i=1}^{N} y_i \log \left(\frac{\pi_i}{1 - \pi_i}\right) + n_i \log (1 - \pi_i) + \log \binom{n_i}{y_i} \tag{1}$$

let $\hat{\pi}$ denote the maximum likelihood estimates for the probabilities of success or failure and let $\hat{y}_i = n_i \hat{\pi}_i$ denote the fitted values. Then the loglikelihood function evaluated at these values is

$$l(\beta, y) = \sum \left[ y_i \log \left(\frac{\hat{y}_i}{n_i - \hat{y}_i}\right) + n_i \log \left(\frac{n_i - \hat{y}_i}{n_i}\right) + \log \left(\binom{n_i}{y_i}\right) \right] \tag{2}$$

**2.4. Generalized linear model:**

We want to describe the proportion of successes, $\pi_i = Y_i/n_i$, in each subgroup in terms of factor levels and other explanatory variables. As $E(Y_i) = n_i \pi_i$ and so $E(P_i) = \pi_i$, The probability function $\pi_i$ known as link function

$$g(\pi_i) = x_i^T \beta \tag{3}$$

**Table 1**: Regression and tolerance model corresponding to the link functions under GLM

| Link type | Regression Model | Tolerance Model |
|---|---|---|
| Logit | $\log \left(\frac{\pi_i}{1 - \pi_i}\right) = x_i^T \beta$ | $\pi_i = \frac{\exp(x_i^T \beta)}{1 + \exp(x_i^T \beta)}$ |

| Probit | $\Phi^{-1}(\pi_i) = x_i^T\beta$ | $\pi_i = \Phi(x_i^T\beta)$ |
| --- | --- | --- |
| Cloglog | $\log(-\log(1-\pi_i)) = x_i^T\beta$ | $\pi_i = \exp(-\exp(x_i^T\beta))$ |
| Cauchy | $t\tan\left(\left(\pi_i - \frac{1}{2}\right)\pi\right) = x_i^T\beta$ | $\pi_i = \frac{1}{\pi}\arctan\left(\frac{x_i^T\beta}{t}\right) + \frac{1}{2}$ |

The probability $\pi_i$ is belongs to [0,1]. The fitted linear model $x^T\beta$ could be a numerical value $(-\infty, \infty)$. The link function restricts the linear model into the interval [0,1], defined as $\pi_i$ known as **probability** cumulative distribution [6].

$$\pi = \int_{-\infty}^{t} f(x)dx \qquad (4)$$

Where $f(x) \geq 0$ and $\int_{-\infty}^{\infty} f(x)dx = 1$. The probability density function f(x) is called the tolerance distribution.

### 2.5. Goodness of fit test:

The measurements of goodness of fit for the classical linear regression are $R^2, RMSE$, while the criteria for the GLM are AIC and residual deviance. Smaller AIC indicates better model.

$$AIC = 2k - 2\ln(\hat{L}) \qquad (5)$$

Where k is the number of estimated parameter, $\hat{L}$ is the maximum value for the maximum likelihood model.

**Table 2**: Cross-table for socio-demographic determinants with response variable

| Variables | $\chi^2$ | P-value |
| --- | --- | --- |
| Antenatal care (ANC visit) | 133.03*** | 0.00 |
| Delivery place | 127.74*** | 0.00 |
| Residential area | 16.447** | 0.036 |

| Wealth quantile | 155.65*** | 0.00 |
| Gender | 33.54*** | 0.00 |
| Mother education | 205.72*** | 0.00 |
| Father education | 119.96*** | 0.00 |
| Child weight at birth | 133.01*** | 0.00 |

Note: '*' 10%, '**' 5%, '***' 1% significant level.

Deviance, $D = 2[l(b_{max}, y) - l(b, y)] = 2\sum \left[ y_i \log \left( \frac{y_i}{\hat{y}_i} \right) + (n_i - y_i) \log \left( \frac{n - y_i}{n - \hat{y}_i} \right) \right]$

Deviance, $D = (0, \infty)$, $D = 0$ indicates perfect model. Increasing the value of $D$ decreases the model acceptability.

## 3. Result & Discussion:

Table 2 shows that the socio-demographic factors are significantly associated with the response variable 'poportion of child malnutrition' under 10% significant level. Table 3 represents the coefficient and standard error of co-variate of binomial regression with link functions. Probit model is considered an appropiate model based on the lowest standard error of covariates under 1% significant level. Having mother antenatal care (ANC visit) during pregnant, probability of child malnutrition is decreased by 0.248 . The chance of children with poor health living in rural area is reduced by 0.104 than children living in urban area. Likelihood of child malnutrition in middle and high class family is fallen off by 0.122 and 0.192 respectively as compared to the children in poor class family. Female children are lower likely (0.120) to becoming a malnutrition as compared to male children. Possibility of child's poor health condition are dropped by 0.181 for primary educated mother, 0.297 for secondary educated mother and 0.309 for higher

Table 3: Coefficents of covariate from binomial regression model with four link functions

| Variable | Labels | Logit | Probit | Cloglog | Cauchy |
|---|---|---|---|---|---|
| Antenatal care (ANC) | Yes | −0.248***(0.054) | -0.154***(0.034) | −0.185**(0.040) | −0.205***(0.045) |
| | No(ref) | - | - | - | - |
| Delivery place | Home | 0.152***(0.045) | 0.093***(0.028) | 0.118***(0.035) | 0.136***(0.040) |
| | Hospital(ref) | - | - | - | - |
| Residential area | Rural | −0.167***(0.056) | -0.104***(0.034) | -0.124***(0.043) | -0.139***(0.050) |
| | Urban(ref) | - | - | - | - |
| Wealth quintile | Rich | −0.311***(0.055) | -0.192***(0.034) | -0.239***(0.044) | -0.263***(0.049) |
| | Middle | −0.195***(0.056) | -0.122***(0.035) | -0.144***(0.044) | -0.154***(0.049) |
| | poor | - | - | - | - |
| Gender | Female | -0.194***(0.040) | -0.120***(0.025) | -0.149***(0.031) | -0.165***(0.035) |
| | Male(ref) | - | - | - | - |
| Mother educstion | Primary | −0.290***(0.076) | -0.181***(0.047) | -0.197***(0.054) | -0.233***(0.063) |
| | Secondary | −0.477***(0.075) | -0.297***(0.047) | -0.345***(0.053) | -0.395***(0.062) |
| | Higher | −0.495***(0.093) | -0.307***(0.058) | -0.367***(0.071) | -0.423***(0.082) |
| | None (ref) | - | - | - | - |
| Father education | Primary | 0.005(0.052) | 0.003(0.032) | 0.004(0.039) | 0.004(0.044) |
| | Secondary | −0.116**(0.056) | −0.072**(0.035) | -0.092**(0.044) | −0.102**(0.049) |
| | Higher | −0.204**(0.083) | −0.124**(0.050) | −0.165**(0.067) | -0.193**(0.079) |
| | None (ref) | - | - | - | - |
| Child birth weight | Overweight | −0.239***(0.067) | -0.146***(0.040) | -0.196***(0.055) | -0.221***(0.064) |
| | Underweight | 0.438***(0.052) | 0.271***(0.032) | 0.325***(0.038) | 0.372***(0.044) |
| | Average(ref) | - | - | - | - |

Note: '*' 10%, '** 5%, '***' 1% significant level.; ref (reference group): 1

Table 4: Goodness of fit test for binomial regression with four link functions

| Statistics | Logit | Probit | Cloglog | Cauchy |
|---|---|---|---|---|
| Deviance | 11696.39 | 11696.35 | 11697.97 | 11697.31 |
| AIC | 12936.27 | 12936.23 | 12937.84 | 12,937.18 |

educated mother; the corresponding likelihood for father education are decreased by 0.072 and 0.124 for secondary education and higher education respectively. Children born in home increases the chance of being malnourishment by 0.093 as compare to children delivered in hospital. Likeliness of child weak health are decreased by 0.146 for overweight during birth, while the chance of child malnutrition is raised by 0.271 for underweight at birth as compared to average weight at birth.

Different socio-demographic factors are significantly associated with child malnutrition in this study. Mother Antenatal care (ANC visit) is statistically associated reducing child undernourishment. Our study is consistent with several cross sectional study. A study in Madhya Pradesh, India, shows that ANC visit may not causes of child poor health [7,16]. But poor care of mother (ANC visit) during pregnancy could have high risk of child health. An evidence of another study in Burkina Faso, a country of West Africa, noted that a statistical association is found between mother ANC visit and child birth-weight.

Available access of mother ANC visit tends to lower low birth weight child [8]. Low birth weight (LBW) is one of the key factors of child malnuorishment. A study in Bangladesh evidences that LBW children are more likely to be affected communicable and non-communicable diseases like respiratory infection, lung disorder, anemia, diarrhea, etc. The children of LBW are growth failure in physically and mentally during early life. Our study supports that LBW is positively associated with child poor health [9].

The prevalence of child malnutrition in rural and urban areas are contradicting over the decades. A study in Bangladesh and Egypt show child malnourishment in rural area are lowered compare to the urban area. Outdoor activity in Urban area are becoming shirnakge day by day. Children and their parents are not involve in the outdoor activities. Besides, industries in urban area are polluting surrounding ecosystem. Consequently child experienced with poor health are prevalent in urban area that support our study [10,11,12].

Gender inequality is commonly shown among newborn children. Study show female children are more likely to be malnourished as compared to male children. Uneducated family may be less concern about health status of female children [13, 14]. But our study contradicts with the statement. Educated parents are treating their new born children as equally. They are more concern about overall well-being of children. Consequently increasing parent education level reduces the child malnutrition in those families [15]. The same scenario is found in our study. Educated family maintain more stable economic status. Economic inequality is considered one of the major burden of child malnutrition. Household economic status has a positive association with reducing child under-nourishment. Parent in the high economic status family are capable of maintaining all types of children needs [14, 17]. Our study also support that higher level of economic status in household level decreases the child malnutrition.

### 4. Conclusion:

Binomial regression model is being using for dichotomous or proportional case of binary response variable. Generally, cloglog link function works for proportional case, while logit and probit use for binary case. But in our work, the probit model shows the minimum standard error of the coefficients. Besides, the goodness of fit test (deviance, AIC) also support the probit model. We need to keep in mind the rule of link functions in the binomial model.

**Acknowledgements:**

We acknowledge UNICEF and BBS for getting free data set 'Multiple Indicator Cluster Survey-2019, Bangladesh'.